\documentclass[aps,twocolumn, superscriptaddress, english, nofootinbib, 10pt]{revtex4-2}

\usepackage{float}

\usepackage{amssymb, amsmath}
\usepackage[utf8]{inputenc}
\usepackage{subfigure}
\usepackage{comment}
\usepackage{graphicx}
\usepackage{hyperref}
\usepackage{enumerate}
\usepackage{lipsum}
\usepackage{xspace}
\usepackage{aas_macros}
\usepackage{enumitem}
\usepackage{booktabs}

\usepackage[usenames,dvipsnames]{xcolor}
\hypersetup{
    colorlinks = true,
    citecolor = {MidnightBlue},
    linkcolor = {BrickRed},
    urlcolor = {BrickRed}
}

\newcommand{\nv}{\hat{\bf n}}

\newcommand{\absum}{{\tt AbacusSummit}\xspace}

\begin{document}

\title{kSZ for everyone: the pseudo-$C_\ell$ approach to stacking}

\author{Lea Harscouet}
\email{lea.harscouet@physics.ox.ac.uk}
\affiliation{Department of Physics, University of Oxford, Denys Wilkinson Building, Keble Road, Oxford OX1 3RH, United Kingdom}
\author{Kevin Wolz}
\email{kevin.wolz@physics.ox.ac.uk}
\affiliation{Department of Physics, University of Oxford, Denys Wilkinson Building, Keble Road, Oxford OX1 3RH, United Kingdom}
\author{Amy Wayland}
\affiliation{Department of Physics, University of Oxford, Denys Wilkinson Building, Keble Road, Oxford OX1 3RH, United Kingdom}
\author{David Alonso}
\affiliation{Department of Physics, University of Oxford, Denys Wilkinson Building, Keble Road, Oxford OX1 3RH, United Kingdom}
\author{Boryana Hadzhiyska}
\affiliation{Institute of Astronomy, Madingley Road, Cambridge, CB3 0HA, UK}
\affiliation{Kavli Institute for Cosmology Cambridge, Madingley Road, Cambridge, CB3 0HA, UK}

\begin{abstract}
  We present a harmonic-space estimator for the cross-correlation between the kinematic Sunyaev-Zel'dovich effect and the reconstructed galaxy momentum field that offers several practical advantages over the traditional stacking approach. The estimator is easy to deploy using relatively modest computational resources and recovers all information available in the galaxy-kSZ cross-correlation. In particular, by using well-understood power spectrum estimation techniques, its statistical uncertainties,  including potential correlated uncertainties with other large-scale structure observables, can be easily and accurately estimated. Moreover, standard kSZ stacking measurements can be reconstructed exactly from the estimator at a lower computational cost, employing harmonic-space, catalog-level techniques to recover all small-scale information.
\end{abstract}

\maketitle

\section{Introduction}\label{sec:intro}
  As the size and sensitivity of new large-scale structure (LSS) and cosmic microwave background (CMB) datasets have grown over the last decade, our measurements have become sensitive to complex astrophysical effects that can no longer be modeled via simplified parametrizations \citep{2020JCAP...04..019S,2011.15018,2404.06098,2506.11943,2507.07892}. Our poor understanding of these processes, particularly on small scales (e.g. the impact of active galactic nuclei (AGN) feedback on the suppression of small-scale matter fluctuations \citep{1905.06082}, or the detailed clustering of galaxies \citep{2101.11014,2101.12187,2307.03226}), requires us to marginalize over astrophysical models with an increasing number of degrees of freedom, leading to a severe degradation in the cosmological constraints that could be recovered otherwise. This, as well as the need to obtain data-driven priors over these unknown astrophysical processes, has pushed the community towards the analysis of canonical cosmological probes (e.g. the clustering of galaxies and weak gravitational lensing measurements) in combination with other LSS tracers with a complementary sensitivity to the main astrophysical processes that dominate our uncertainties \citep{2109.04458,2308.01856,2309.11129,2401.18072,2404.06098,2411.04088,2412.12081,2505.20413,2509.10455,2510.25596}.

  The kinematic Sunyaev-Zel'dovich effect (kSZ) is a perfect example of this paradigm \citep{astro-ph/9808050,1811.02310}. The inverse-Compton scattering of CMB photons by free electrons in the intergalactic medium (IGM) causes an additional contribution to the CMB anisotropies. While the thermal motion of these electrons modifies the CMB spectrum, in the so-called ``thermal'' Sunyaev-Zel'dovich effect (tSZ), the kSZ is sourced by the bulk peculiar motion of the structures in which these electrons reside, which leaves the CMB spectrum unchanged. Crucially, the kSZ effect is directly sensitive to the density of ionized gas in the IGM, as well as the peculiar velocity field. As a direct probe of gas, the kSZ can provide a direct measurement of the distribution of baryons, including the impact of AGN feedback, and thus serve as an external calibrator for baryonic effects \citep{2404.06098,2407.07152,2410.19905,2503.19870,2506.17379,2510.15822}. This is typically achieved by measuring the amplitude of the kSZ effect around galaxies, in the so-called ``kSZ stacking'' approach \citep{liMatchedFilterOptimization2014,1510.06442,2009.05557,2305.06792,2407.07152,2503.19870}: CMB maps, filtered to suppress the large-scale contribution from the uncorrelated primary CMB fluctuations, are stacked at the positions of sources weighted by an estimate of the galaxy velocities. These kSZ stacking measurements have provided evidence for the strength of AGN feedback to be larger than previously expected or predicted by hydrodynamical simulations, leading to a stronger suppression in the small-scale matter power spectrum as measured in cosmic shear analyses \citep{2009.05557,2404.06098,2407.07152,2410.19905,2512.02954}.

  To ensure that uncertainties in astrophysical processes are robustly propagated in cosmological inference, and self-calibrated by including external measurements (e.g. kSZ in the context of baryonic effects), an ideal analysis would combine these measurements with the canonical cosmological probes, and analyze them simultaneously within a unified physics-based model (e.g. one linking the stacked kSZ signal and the power spectrum suppression), taking into account all observational effects and cross-probe correlations \citep{1010.0744,1909.02179,2006.00008}. Such joint analyses also improve in robustness when all different probes are processed under a common framework, making it possible to apply self-consistent analysis choices (e.g. scale cuts for different tracers), enabling the use of a coherent error model for all measurements, including correlated uncertainties, and simplifying the production of theoretical predictions for all probes under a common scheme. In this sense, incorporating the standard kSZ stacking measurements into cosmological analyses poses a number of practical challenges: accounting for interpolation and pixelization effects inherent to stacking, the impact on scale cuts of the filtering used, the development of a robust covariance estimate, masking and survey geometry effects, and the relatively high computational cost of real-space correlation estimators \citep{astro-ph/0307393,1210.1833,2020MNRAS.491.3022S}. While all these challenges can be addressed (e.g. using simulation-based covariances, forward-modeling pixelization effects, avoiding masking effects via catalog-level cuts), doing so increases the technical complexity of the analysis pipelines, and the level of expertise needed to reliably incorporate kSZ measurements in any given study.

  The integration of kSZ measurements into the widely-used framework of correlation functions is not out of reach, however. Many traditional, real-space kSZ measurements are in fact indirect probes of the galaxy-galaxy-temperature ($ggT$) bispectrum; \cite{smithKSZTomographyBispectrum2018} establishes a direct equivalence between $ggT$ and -- among other estimators -- the pairwise velocity method of \cite{handEvidenceGalaxyCluster2012}, which led to the first detection of the kSZ effect, and the stacking estimator of \cite{liMatchedFilterOptimization2014,1510.06442}, described above. Other works have also measured the kSZ effect via the galaxy-temperature-temperature $gTT$ bispectrum (recently in \cite{patkiNovelBispectrumEstimator2024} for projected fields) or its compressed proxy, the $C_\ell^{T^2 g}$ skew spectrum in \cite{hillKinematicSunyaevZeldovichEffect2016}. Our endeavour is therefore part of a wider effort to turn long-established kSZ detection methods into reliable tools that can be used in concordance with current methods for large-scale structure studies.

  In this paper we present an angular power spectrum approach to measure the kSZ signal around galaxies which, as we will show, is at least equivalent to the kSZ stacking estimator in terms of information content, while resolving the practical challenges listed above. This new estimator, $C_\ell^{\pi T}$, probes the $ggT$ bispectrum by cross-correlating the momentum field $\pi_g$, built from the galaxy density and reconstructed velocity fields, and the CMB temperature field $T$ (an approach originally advocated in \cite{0903.2845}).
  Using well-tested methods developed in the context of pseudo-$C_\ell$ estimation \citep{astro-ph/0105302,1809.09603,2407.21013}, the resulting power spectrum measurements can be obtained at a low computational cost, are robust against masking effects, and their statistical uncertainties can be accurately and easily estimated. The estimator may be seamlessly combined with power spectrum measurements involving other LSS tracers using the same framework, easily enabling the multi-tracer analysis needed for a robust self-calibration of astrophysical effects. As we were preparing this manuscript, we learnt of similar work presented in upcoming publications \cite{quinprep,hadzhiyskainprep}, where this approach is applied to new datasets. Other works have also recently begun to model and study the kSZ signal around galaxies in terms of the harmonic-space spectrum \citep{2509.18732,2511.20595}.
  
  This paper is structured as follows. Section \ref{sec:meth} introduces the kSZ-galaxy power spectrum estimator, and describes its advantages and its relation with kSZ stacking. Section \ref{sec:res} validates our implementation of this framework, demonstrates its ability to recover the stacking measurements at the estimator level, and showcases its application to real data. We then conclude in Section \ref{sec:conc}.

\section{Methods}\label{sec:meth}

  \subsection{kSZ angular power spectra}\label{ssec:meth.cl}
  
    \subsubsection{Galaxy momentum maps}\label{sssec:meth.cl.map}
    
      The kSZ contribution to the temperature fluctuations is
      \begin{align}\label{eq:DeltaT_kSZ}
        &\Delta T_{\rm kSZ}(\nv)=\int d\chi\,q_{\rm kSZ}(\chi)\,[1+\delta_e(\chi\nv,z)]\,\nv\cdot{\bf v}(\chi\nv,z),\\
        &q_{\rm kSZ}(\chi)\equiv T_{\rm CMB}\sigma_T\bar{n}_{e,c}(1+z)^2,
      \end{align}
      where $\bar{n}_{e,c}$ is the comoving mean number density of electrons, $T_{\rm CMB}$ is the CMB temperature, and $\sigma_T$ is the Thomson scattering cross section. $\delta_e$ is the electron overdensity, and ${\bf v}$ is the peculiar velocity field, in units where the speed of light is $c=1$. Redshifts $z$ and comoving distances $\chi$ are implicitly related to one another on the lightcone. The kSZ effect is therefore a line-of-sight projection of the radial ``electron momentum density'' $q_e\equiv (1+\delta_e)\,\nv\cdot{\bf v}$,\footnote{Note that in this definition, the momentum density has units of velocity, being normalized by the mean number density.} and is sensitive to the large-scale structure both through the density and velocity fields.

      Given a sample of galaxies, we may attempt a tomographic measurement of the kSZ effect by first constructing a proxy for the corresponding projected ``galaxy momentum density'' using the galaxy information, before correlating it with $\Delta T_{\rm kSZ}$. Since galaxies themselves trace the large-scale structure, this may be achieved by simply adding the radial velocities of all galaxies located in a given sky patch. That is, for a given sky pixel $p$, we may construct the following quantity:
      \begin{equation} \label{eq:mom_pix}
        \pi_{g,p}\equiv\frac{1}{\bar{N}_p}\sum_{i\in p}w_i\nv_i\cdot{\bf v}^g_i,
      \end{equation}
      where the sum is over all objects with coordinates $\nv_i$ lying inside pixel $p$, $w_i$ are the associated galaxy weights, and ${\bf v}^g_i$ is an estimate of their velocity. $\bar{N}_p$ is an estimate of the expected mean number of galaxies that could have been observed in pixel $p$, and is included in order to account for potential variations in survey completeness and pixel area. $\pi_g$ thus defined corresponds to the quantity
      \begin{equation}
        \pi_g(\nv)\equiv \int dz\,p(z)\,[1+\delta_g(\chi\nv,z)]\,\nv\cdot{\bf v}^g(\chi\nv,z),
      \end{equation}
      where $\delta_g$ is the galaxy overdensity, and $p(z)$ is the redshift distribution of the sample.

      Note that we purposely denote the galaxy velocities ${\bf v}^g$, distinct from the true peculiar velocity field ${\bf v}$. This is because it is rarely possible to estimate the true galaxy velocities and, instead, it is most common to employ reconstruction techniques to estimate ${\bf v}^g$ from $\delta_g$. For our purposes here, we will simply assume that an estimate of ${\bf v}^g_i$ exists for each galaxy, and its origin will be otherwise irrelevant.

      Once $\pi_g$ has been constructed, our final aim is to estimate the cross-spectrum:
      \begin{equation} \label{eq:cl_piT}
        C_\ell^{\pi T}\equiv \left\langle \Delta T_{{\rm kSZ}, \ell m}\,\pi_{g,\ell m}^*\right\rangle.
      \end{equation}
      We can do this following the standard pseudo-$C_\ell$ estimator \citep{astro-ph/0105302,1809.09603}. In this case, the cross-power spectrum between two fields $a$ and $b$ is calculated assuming a diagonal pixel-space inverse covariance matrix for each field, defined by their masks. In most cases we will aim to extract $\Delta T_{\rm kSZ}$ from either a component-separated CMB$+$kSZ map or from a simple frequency map, under the assumption that, out of all Galactic and extragalactic components contributing to it, the only one that correlates with the velocity-weighted density field is the kSZ. Either way, this map will normally be accompanied by its own mask $w_{\rm kSZ}(\nv)$, typically constructed through a combination of point-source and diffuse Galactic masks in addition to any mask associated with the actual survey footprint of the data. In contrast, a typical choice for the mask for $\pi_g$ is $\bar{n}_p\equiv \bar{N}_p \Omega_p^{-1}$, the expected 2D-projected number density of galaxies in each pixel (with the dimension of inverse solid angle), where $\Omega_p$ is the pixel area. 
      For noise-dominated data, and assuming that the velocity reconstruction uncertainties are homogeneous across the survey footprint, this choice is close to an inverse-variance weighting of the field\footnote{For simplicity, we assume that $\bar{n}_p$ is the mask of choice in this work. This would correspond to the inverse-variance map of an inhomogeneous Poisson-sampled galaxy field. If an estimate of the spatially-varying statistical uncertainties in the reconstructed velocities is available, this could be incorporated into a more optimal mask.}. In this case, the masked galaxy momentum field is \
      \begin{equation}
        \tilde{\pi}_{g,p}=\frac{1}{\Omega_p}\sum_{i\in p}w_i v^g_{r,i},
      \end{equation}
      where $v^g_{r,i}\equiv \nv_i\cdot{\bf v}^g_i$.

      In most cases, $\bar{n}_p$ will be constructed from a continuous map, $m^g_p$, tracing the survey's completeness across the sky. In this case, $\bar{n}_p$ can be calculated by requiring that its integral across the sky should be the total (weighted) number of galaxies:
      \begin{equation}
        \bar{n}_p=\alpha_N\,m^g_p\, ,\hspace{6pt}\alpha_N\equiv\frac{\sum_{i\in D}w_i}{\sum_{p'} m^g_{p'} \Omega_{p'}},
      \end{equation}
      where the sums in $\alpha_N$ run over all galaxies in the sample and all pixels in the sky. Another common alternative is to define the survey completeness in terms of a random catalog, with the completeness at a given point in the sky given by the local weighted number density of random objects. In this case $\bar{n}_p$ may be constructed as
      \begin{equation}
        \bar{n}_p=\frac{\alpha_r}{\Omega_p}\,\sum_{i\in p}w^r_i,\hspace{12pt}
        \alpha_r=\frac{\sum_{j\in D}w_j}{\sum_{k\in R}w^r_k},
      \end{equation}
      where $\sum_{i\in p}$ involves summing over all random objects in pixel $p$, with weights $w^r_i$, and the sums $\sum_{i\in D}$ and $\sum_{i\in R}$ run over all data and random objects, respectively.

      With the masked fields and associated masks for both fields $\Delta T_{\rm kSZ}$ and $\pi_g$ constructed as we have just described, the pseudo-$C_\ell$ estimate of their cross-spectrum can be extracted efficiently following the steps outlined in e.g. \cite{1809.09603}. Moreover, its covariance matrix may also be estimated accurately and easily using the so-called improved narrow-kernel approximation (iNKA) as described in \cite{1906.11765,2010.09717}, using the measured pseudo-spectra of the fields involved as input.
      
    \subsubsection{Avoiding pixels: catalog-based $C_\ell$s}\label{sssec:meth.cl.cat}
      As described in \cite{2312.12285,2407.21013}, the galaxy density is ultimately determined by the positions of the galaxies themselves, and traditionally the only reason to bin them into some form of regular pixel grid, is the requirement of most numerical spherical harmonic transform (SHT) routines to use such a grid to achieve both efficiency and accuracy. However, the recent development of SHT algorithms on irregularly-shaped grids \cite{2304.10431} has made it possible to estimate the power spectra of fields sampled at the positions of discrete sources without pixelization. Since much of the kSZ signal resides at small angular scales ($\ell\gtrsim3000$, the larger scales often being dominated by primary CMB ``noise''), we can take advantage of these methods to efficiently recover the kSZ-galaxy power spectrum at these scales.

      For example, the observed number density of galaxies, and its spherical harmonic coefficients, are given by
      \begin{align}
        &n_g(\nv)=\sum_{i\in D}w_i\,\delta^D(\nv,\nv_i),\\
        &n_{g,\ell m }=\sum_{i\in D}w_i\,Y_{\ell m}^*(\nv_i),
      \end{align}
      where $\delta^D(\nv,\nv_i)$ is the Dirac delta function on the sphere. As in the previous section, let us assume that a map of the expected number density of galaxies, $\bar{n}(\nv)$, exists. Then, following this formalism, we may express the masked galaxy momentum field (and its SHT) as
      \begin{align}\label{eq:pi_cat}
        &\tilde{\pi}_g(\nv)=\sum_{i\in D}w_iv^g_{r,i}\,\delta^D(\nv,\nv_i),\\\label{eq:pilm_cat}
        &\tilde{\pi}_{g,\ell m}=\sum_{i\in D}w_iv^g_{r,i}\,Y^*_{\ell m}(\nv_i),
      \end{align}
      assuming $\bar{n}$ to be its mask.

      The only other ingredient needed to estimate the power spectra of $\pi_g$ is the spherical coefficients of the mask $\bar{n}_{\ell m}$. Again, as in the previous section, $\bar{n}$ may be provided either in terms of a completeness map $m^g$, or through a random catalog. In the first case, $\bar{n}$ is simply given by $\bar{n}_{\ell m}=\tilde{\alpha}_n\,m^g_{\ell m}$, where
      \begin{equation}\label{eq:alpha_n}
        \tilde{\alpha}_n\equiv\frac{\sum_{i\in D}w_i}{\int d\nv\,m^g(\nv)}.
      \end{equation}
      In the case of a random catalog, we may again take advantage of irregular-grid SHTs to calculate $\bar{n}_{\ell m}$
      \begin{equation}
        \bar{n}_{\ell m}=\alpha_r\sum_{i\in R}w_i^r\,Y^*_{\ell m}(\nv_i),
      \end{equation}
      avoiding any finite-pixel effects.

      The power spectra of two discretely-sampled fields $\{a,\,b\}$, with masks $\{v,w\}$, can then be estimated following the standard pseudo-$C_\ell$ estimator \citep{1809.09603,2407.21013}:
      \begin{enumerate}
        \item Let $\tilde{a}\equiv v\,a$ and $\tilde{b}\equiv w\,b$ be the masked fields. First, estimate their pseudo-$C_\ell$, defined as the na\"ive $C_\ell$ estimator ignoring the impact of masking:
        \begin{equation}\label{eq:pcl}
          \tilde{C}^{ab}_\ell\equiv\frac{1}{2\ell+1}\sum_m\tilde{a}_{\ell m}\tilde{b}^*_{\ell m}.
        \end{equation}
        Since masking leads to mode-coupling, $\tilde{C}^{ab}_\ell$ is also sometimes called the ``mode-coupled'' $C_\ell$.
        \item Calculate the pseudo-$C_\ell$ of the masks, $\tilde{C}^{vw}_\ell$ and, from it, the mode-coupling matrix (MCM) $M^{vw}_{\ell\ell'}$, encoding the statistical coupling between angular multipoles induced by masking, and relating the ensemble average of $\tilde{C}^{ab}_\ell$ with the true underlying power spectrum of the fields.
        \begin{equation}\label{eq:mcm}
          \left\langle\tilde{C}^{ab}_\ell\right\rangle=\sum_{\ell'}M^{vw}_{\ell\ell'}C^{ab}_\ell.
        \end{equation}
        \item Invert the MCM to obtain an unbiased estimate of $C_\ell^{ab}$. This step is often preceded by binning the power spectrum in $\ell$ into so-called ``bandpowers'', which avoids numerical inversion issues when the masks are complex or cover a small portion of the sky, and shortens the final data vector if a high resolution in $\ell$ is not necessary.
      \end{enumerate}
      In addition to these steps, one caveat must be borne in mind in the case of auto-correlations of catalog-based fields where, as noted in \cite{2407.21013,2408.16903}, the shot-noise contributions to $\tilde{C}^{ab}_\ell$ and $\tilde{C}^{vw}_\ell$ must be subtracted to avoid aliasing effects, simultaneously hardening the estimator against any white noise contribution. This subtlety is of less importance in the case of $C_\ell^{\pi T}$, since we will rarely want to calculate the auto-correlation of $\pi_g$.

      Finally, since this will be instructive for our discussion in the next section, let us consider the particular case in which the temperature map entering $C_\ell^{\pi T}$ is masked with a relatively simple binary mask $w_T$ that encompasses the footprint over which $\pi_g$ has been measured (i.e. $w_T\,\bar{n}=\bar{n}$). As we will see in the next section, this is in fact a common practice in kSZ stacking studies. Approximating $C_\ell^{\pi T}$ to be a slowly-varying function, and that $M^{w_T\bar{n}}_{\ell\ell'}$ is sharply peaked around $\ell\simeq\ell'$, we can approximate Eq. \ref{eq:mcm} as
      \begin{equation}
        \left\langle \tilde{C}^{\pi T}_\ell\right\rangle\simeq
        C_\ell^{\pi T}\sum_{\ell'}M^{w_T\bar{n}}_{\ell\ell'}=\langle w_T\bar{n}\rangle_{\rm sky}C_\ell^{\pi T},
      \end{equation}
      where $\langle\cdot\rangle_{\rm sky}$ denotes averaging over the sky, and in the second equality we used a well-known property of the MCM (see e.g. \cite{astro-ph/0105302,2010.09717}). The prefactor in the final expression is then
      \begin{equation}
        \langle w_T\bar{n}\rangle_{\rm sky}=\langle \bar{n}\rangle_{\rm sky}=\int\frac{d\nv}{4\pi}\bar{n}(\nv)=\frac{\sum_{i\in D}w_i}{4\pi}.
      \end{equation}
      Thus, we may define the following approximate estimator of the kSZ-galaxy power spectrum:
      \begin{equation}\label{eq:defcalC}
        {\cal C}_\ell^{\pi T}\equiv\frac{\tilde{C}^{\pi T}_\ell}{\langle\bar{n}\rangle_{\rm sky}}.
      \end{equation}
      The factor $1/\langle \bar{n}\rangle_{\rm sky}$ corrects the overall amplitude of $\tilde{C}^{\pi T}_\ell$ for the effects of masking, although no attempt is made to account for mode-coupling in ${\cal C}^{\pi T}_\ell$.

  \subsection{The connection with kSZ stacking}\label{ssec:meth.stack}
    The more standard method to extract the kSZ-galaxy cross-correlation signal used in the literature \citep{1510.06442,2009.05557,2305.06792,2407.07152,2503.19870} is the so-called ``kSZ stacking'' approach, in which a suitably-filtered CMB  map is stacked around the positions of galaxies weighted by their reconstructed velocities. The aim of this section is to show that the resulting kSZ stacking estimator can be recovered from the kSZ-galaxy angular power spectrum described in the previous section \emph{at the estimator level}: i.e. all of the information recovered by the kSZ stack is encoded in the power spectrum measurements, and can be recovered from it if so desired.

    Let us start by describing the stacking estimator. Specifically, we will follow the prescription of \cite{2009.05557}, employing an aperture photometry filter to reduce the impact of the large-scale primary CMB noise. Consider a galaxy at position $\nv_i$, and let $T(\nv)$ be a map of the CMB temperature fluctuations. We define the aperture-photometry-filtered CMB map on an angular scale $\theta_d$ around this galaxy as:
    \begin{equation}\label{eq:apfilter}
      {\cal T}(\nv_i,\theta_d)=\int\,d\nv\,\tilde{T}(\nv)\,W_{\rm AP}(\nv_i\cdot\nv,\theta_d),
    \end{equation}
    where the compensated aperture photometry (CAP) filter is
    \begin{equation}\label{eq:cap_real}
      W_{\rm AP}(\cos\theta,\theta_d)=\left\{
      \begin{array}{ll}
        1 & \theta \leq \theta_d\\
        -1 & \theta_d< \theta\leq \sqrt{2}\theta_d\\
        0 & \theta > \sqrt{2}\theta_d
      \end{array}
      \right.,
    \end{equation}
    and, in Eq. \ref{eq:apfilter} we have added a tilde over $T$ to denote that it is a masked CMB map. The CAP filter is designed to suppress the impact of large-scale ($\ell\lesssim100$) CMB fluctuations in the measurements, by estimating the largest-scale mode of the field in the inner disk of radius $\theta_d$ as the mean in an outer ring of equal area. The kSZ stacking estimator is then defined as the velocity-weighted average of ${\cal T}_i$ over a sample of galaxies:
    \begin{equation}\label{eq:stack_def}
      \Delta\hat{T}_{\rm kSZ}(\theta_d)\equiv Q\sum_{i\in D} w_i\,v_i\,{\cal T}_i(\nv_i,\theta_d),
    \end{equation}
    where $Q$ is a normalization factor. A typical choice for $Q$ is
    \begin{equation}\label{eq:defQ}
      Q=\frac{1}{r_v\,v_{\rm rms}}\frac{1}{\sum_{i\in D}w_i},
    \end{equation}
    where $v_{\rm rms}^2\equiv \sum_{i\in D}w_i(v^g_{r,i})^2/\sum_{i\in D}w_i$ is the variance of the reconstructed velocities. $r_v$ is the correlation coefficient between the reconstructed and true velocities, and is usually included in $Q$ to account for the mismatch between both. It is common in kSZ stacking analyses to only include in the estimator galaxies that are located far enough from the edges of the mask defining the $T$ field to fit the largest disk-ring pair defining the CAP filters used in the measurement. We will assume this to be the case in what follows.

    To rewrite the kSZ stacking estimator in terms of the masked galaxy momentum field we introduced in Eq. \ref{eq:pi_cat}, we insert a Dirac delta function for every object entering the sum in Eq. \ref{eq:stack_def} and integrate over the sphere:
    \begin{align}
      \Delta\hat{T}_{\rm kSZ}(\theta_d)
      &=Q\int d\nv\,\tilde{\pi}_g(\nv)\,{\cal T}(\nv, \theta_d)\\\nonumber
      &=Q\int d\nv_1\,d\nv_2\,\,\tilde{\pi}_g(\nv_1)\,\tilde{T}(\nv_2)\,W_{\rm AP}(\mu_{12}, \theta_d),
    \end{align}
    where $\mu_{12}\equiv\nv_1\cdot\nv_2$ and, in the second line, we have made use of Eq. \ref{eq:apfilter}. We can now expand $\tilde{\pi}_g$ and $\tilde{T}$ in terms of their spherical harmonic coefficients
    \begin{equation}\nonumber
      \tilde{T}(\nv)=\sum_{\ell m}\tilde{T}_{\ell m}Y_{\ell m}(\nv),\hspace{12pt}
      \tilde{\pi}_g(\nv)=\sum_{\ell m}\tilde{\pi}_{g,\ell m}Y_{\ell m}(\nv),
    \end{equation}
    and do likewise for the the aperture-photometry filter
    \begin{align}\nonumber
      W_{\rm AP}(\mu_{12},\theta_d)
      &=\sum_\ell\frac{2\ell+1}{4\pi}W_\ell(\theta_d)\,P_\ell(\mu_{12}),\\
      &=\sum_{\ell m}W_\ell(\theta_d)\,Y_{\ell m}^*(\nv_1)Y_{\ell m}(\nv_2),
    \end{align}
    where $W_\ell$ is the harmonic-space CAP filter, discussed at the end of this subsection, $P_\ell(\mu)$ are the Legendre polynomials, and in the second line we have used the addition theorem for spherical harmonics. Finally, using the orthogonality of the spherical harmonic coefficients, we find
    \begin{equation}
      \Delta\hat{T}_{\rm kSZ}(\theta_d)=Q\sum_\ell (2\ell+1)W_\ell(\theta_d)\,\tilde{C}^{\pi T}_\ell,
    \end{equation}
    where, as in Eq. \ref{eq:pcl}, we have defined
    \begin{equation}
      \tilde{C}^{\pi T}_\ell\equiv\frac{1}{2\ell+1}\sum_m \tilde{\pi}_{g,\ell m}\,\tilde{T}^*_{\ell m}.
    \end{equation}
    Finally, using the definition of $Q$ (Eq. \ref{eq:defQ}) and ${\cal C}^{\pi T}_\ell$ (Eq. \ref{eq:defcalC}), we may rewrite our last result in a more appealing form:
    \begin{equation}\label{eq:cl2stack}
      \Delta\hat{T}_{\rm kSZ}(\theta_d)=\frac{1}{r_vv_{\rm rms}}\sum_\ell\frac{2\ell+1}{4\pi}W_\ell(\theta_d)\,{\cal C}^{\pi T}_\ell.
    \end{equation}
    
    \begin{figure}
        \centering
        \includegraphics[width=0.45\textwidth]{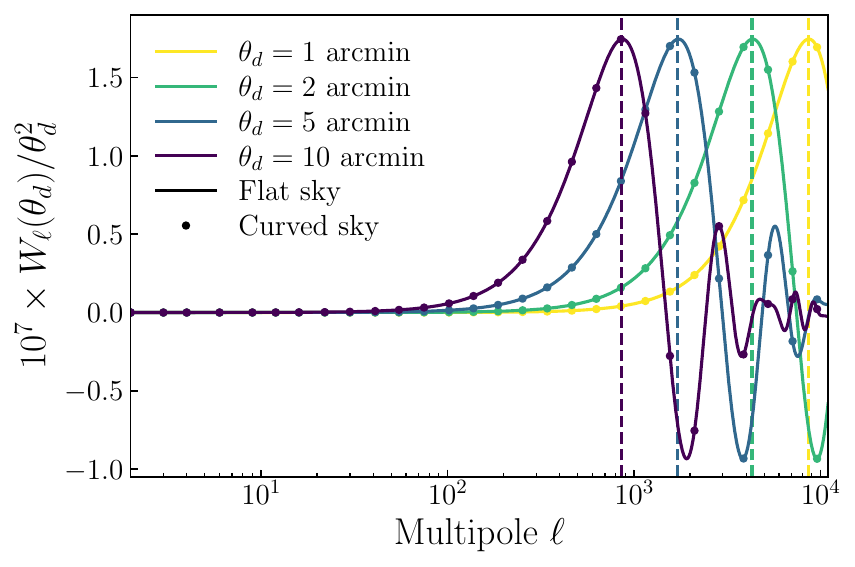}
        \caption{Harmonic transform of the compensated aperture photometry filter for different apertures $\theta_d$. The solid lines show the calculation using the flat-sky approximation (Eq. \ref{eq:cap_fourier}), with the points showing the exact calculation (Eq. \ref{eq:cap_harm}). The flat-sky estimate is accurate enough on all relevant scales, and significantly faster to calculate. The CAP filters peak at a scale $\ell_{\rm CAP}\sim2.5/\theta_d$, with an amplitude proportional to $\theta_d^2$.}
        \label{fig:cap_filters}
    \end{figure}

    The relation between the velocity-weighted kSZ stack and the kSZ-galaxy power spectrum is thus very similar to that existing between the latter and the real-space correlation function of two scalar fields, with Legendre polynomials replaced by the harmonic-space CAP window functions $W_\ell(\theta_d)$. The latter are defined as
    \begin{align}
      W_\ell(\theta_d)
      &=2\pi\int_{-1}^1d\mu\,P_\ell(\mu)W_{\rm AP}(\mu,\,\theta_d)\\
      &\simeq 2\pi\int_0^\infty d\theta\,\theta\,J_0(\ell\theta)\,W_{\rm AP}(\cos\theta, \,\theta_d),
    \end{align}
    where, in the second line, we have made use of the flat-sky approximation (appropriate for the small scales, $\ell\gtrsim 500$, over which the kSZ is normally studied) and $J_\alpha(x)$ is the cylindrical Bessel function of order $\alpha$. Given the simple form of the CAP filter (Eq. \ref{eq:cap_real}), $W_\ell(\theta_d)$ can be calculated analytically:
    \begin{align}\label{eq:cap_harm}
      W_\ell(\theta_d)
      &=\frac{2\pi}{2\ell+1}\left[2\Delta P_\ell(\mu_d)-\Delta P_\ell(\mu_r)\right]\\\label{eq:cap_fourier}
      &\simeq \frac{2\pi}{\ell^2}\ell\theta_d\left[2J_1(\ell\theta_d)-\sqrt{2}J_1(\sqrt{2}\ell\theta_d)\right],
    \end{align}
    where we have defined
    \begin{align}
      &\Delta P_\ell(\mu)\equiv\frac{1}{2}\left(P_{\ell-1}(\mu)-P_{\ell+1}(\mu)\right),\\
      &\mu_d\equiv\cos\theta_d,\hspace{12pt}\mu_r\equiv2\mu_d-1.
    \end{align}
    and, in Eq. \ref{eq:cap_fourier} we have again adopted the flat-sky approximation. As shown in Fig. \ref{fig:cap_filters}, the latter is sufficiently accurate on all relevant scales. The CAP filter peaks at $\ell_{\rm CAP}\sim2.5/\theta_d$, suppressing all large-scale modes (i.e. those dominated by the CMB primary fluctuations) below $\sim\ell_{\rm CAP}$, and oscillates above $\ell_{\rm CAP}$ with an amplitude that decays like $\propto\ell^{-3/2}$.

  \subsection{Advantages of the $C_\ell$ approach}\label{ssec:meth.adv}
    As shown in the previous section, the kSZ stacking estimator can be reconstructed from the kSZ-galaxy power spectrum \emph{at the estimator level}. That is, all the information encoded in the stack can be exactly recovered from $C_\ell^{\pi T}$ by convolving it with the CAP filters following Eq. \ref{eq:cl2stack}. Moreover, since the CAP filters are not a complete set of orthogonal basis functions, Eq. \ref{eq:cl2stack} selects only some of the modes in $C_\ell^{\pi T}$, and thus necessarily incurs some information loss. We will see that this is likely a small fraction, depending on the angular scales included and the impact of the primary CMB anisotropies as a source of irreducible noise. Nevertheless, $C_\ell^{\pi T}$ encodes all information recoverable from the $\pi_g$ and $T$ maps at the level of two-point statistics, making it an optimal observable to use.

    Furthermore, the power spectrum estimator is fast, with the computational cost of spherical harmonic transforms scaling like $\propto\ell_{\rm max}^3$. In contrast, stacking is effectively a two-point correlation function, which, for a given maximum $\theta_d$, scales like $\propto N_{\rm gal}N_{\rm pix}$, where $N_{\rm gal}$ and $N_{\rm pix}$ are the number of galaxies and the number of sky pixels in a square that fits the largest ring for each galaxy, respectively. As we report in Section \ref{ssec:res.stack}, the power spectrum approach is significantly faster (by a factor of $\sim30$ for $\sim 3\times 10^6$ objects) than the stacking approach employing similar computational resources.

    Further advantages are offered by the extensive infrastructure developed for power spectrum estimation using the pseudo-$C_\ell$ approach. First, the estimator is able to account for all masking effects exactly, and therefore avoids the need to discard galaxies close to the mask edges (as commonly done in stacking studies), potentially increasing the sensitivity of the measurements. Additionally, highly accurate and efficient methods have been developed and tested for the calculation of power spectrum covariances \cite{1906.11765,2010.09717}, allowing us to compute the statistical uncertainties of $C_\ell^{\pi T}$ in the approximation of Gaussian fields, without the need to rely on potentially inaccurate bootstrap/jackknife estimates or large numbers of costly mock realizations. This also enables us to easily calculate the correlated uncertainties of $C_\ell^{\pi T}$ with any other power spectrum of cosmological interest. This would, for example, enable a joint analysis of kSZ and $3\times2$-point data from overlapping experiments, ensuring that all correlated uncertainties are accurately taken into account. Moreover, by exploiting statistical isotropy, the power spectra measured on different angular multipoles are largely uncorrelated, and therefore the covariance matrix of $C_\ell^{\pi T}$ is close to diagonal and numerically stable. This is in contrast to the stacking estimator where, due to the ``cumulative'' nature of the CAP filters, measurements on different angular scales $\theta_d$ are highly correlated. This also allows for a quick assessment of the detection significance or the goodness of fit of power spectrum measurements (i.e. performing ``$\chi^2$ by eye'').

    Finally, it is often the case that theoretical models for cosmological observables are naturally expressed in terms of power spectra. For instance, the galaxy-kSZ correlation (either in terms of stacking or $C_\ell^{\pi T}$) is largely determined by the cross-power spectrum between the galaxy and electron overdensities $P_{ge}(k)$ \cite{2509.18732}. Thus, the use of $C_\ell^{\pi T}$ may also be convenient when comparing against theoretical predictions, removing the need for Hankel transforms (e.g. Eq. \ref{eq:cl2stack}) connecting harmonic-space and real-space quantities, or enabling a quick assessment of the relevance of different contributions to the measurements (e.g. 1-halo and 2-halo terms in a halo model calculation of $P_{ge}(k)$).

\section{Results}\label{sec:res}
  \subsection{Implementation and validation}\label{ssec:res.val}
    \begin{figure}
      \centering
      \includegraphics[width=\columnwidth]{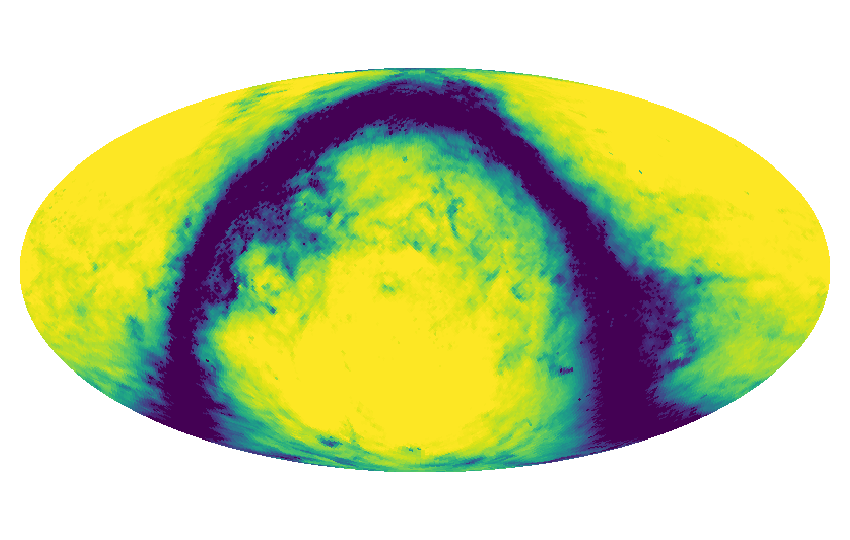}
      \caption{Completeness mask used for the mock survey described in Section \ref{ssec:res.val}.}
      \label{fig:completeness_mask}
    \end{figure}
    We have implemented the general methodology described in the previous section to calculate the power spectrum between galaxy momentum fields and maps in the public pseudo-$C_\ell$ code {\tt NaMaster}\footnote{\url{https://github.com/LSSTDESC/NaMaster}. An example notebook showcasing the kSZ implementation presented here can be found in \url{https://github.com/LSSTDESC/NaMaster/blob/master/doc/kSZ_tutorial.ipynb}.} \cite{1809.09603}. We validate this implementation here using a toy model.

    \begin{figure}
      \centering
      \includegraphics[width=0.98\columnwidth]{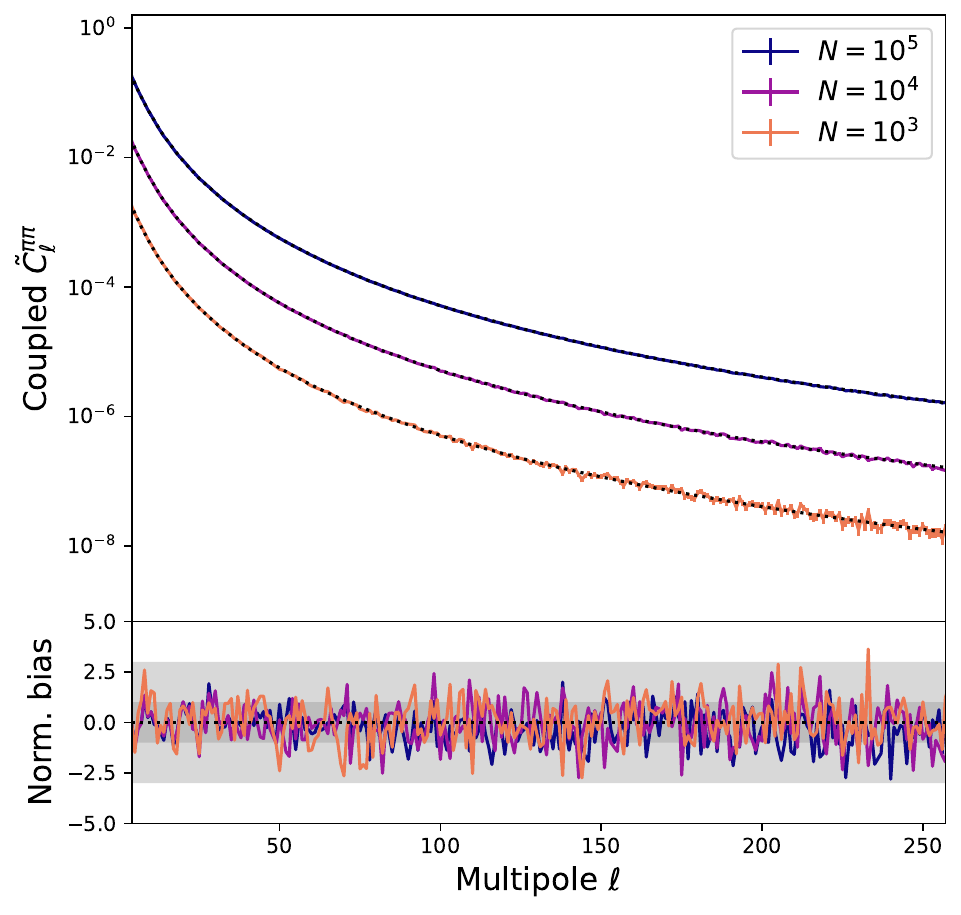}
      \caption{Coupled cross-spectrum between a momentum field computed from a simulated source catalog and its smooth map-based counterpart at $N_{\rm side}=128$, for catalog sizes of $10^5$, $10^4$, $10^3$. We compare against the theoretical input computed from the ground truth coupled with the catalog-based mode-coupling matrix. \textit{Upper panel:} mean over 1000 simulations with bars showing the expected standard error on the mean. \textit{Lower panel:} relative bias, in units of the error on the mean.}
      \label{fig:cl_pi_coupled_sims}
    \end{figure}

    \begin{figure}
      \centering
      \includegraphics[width=\columnwidth]{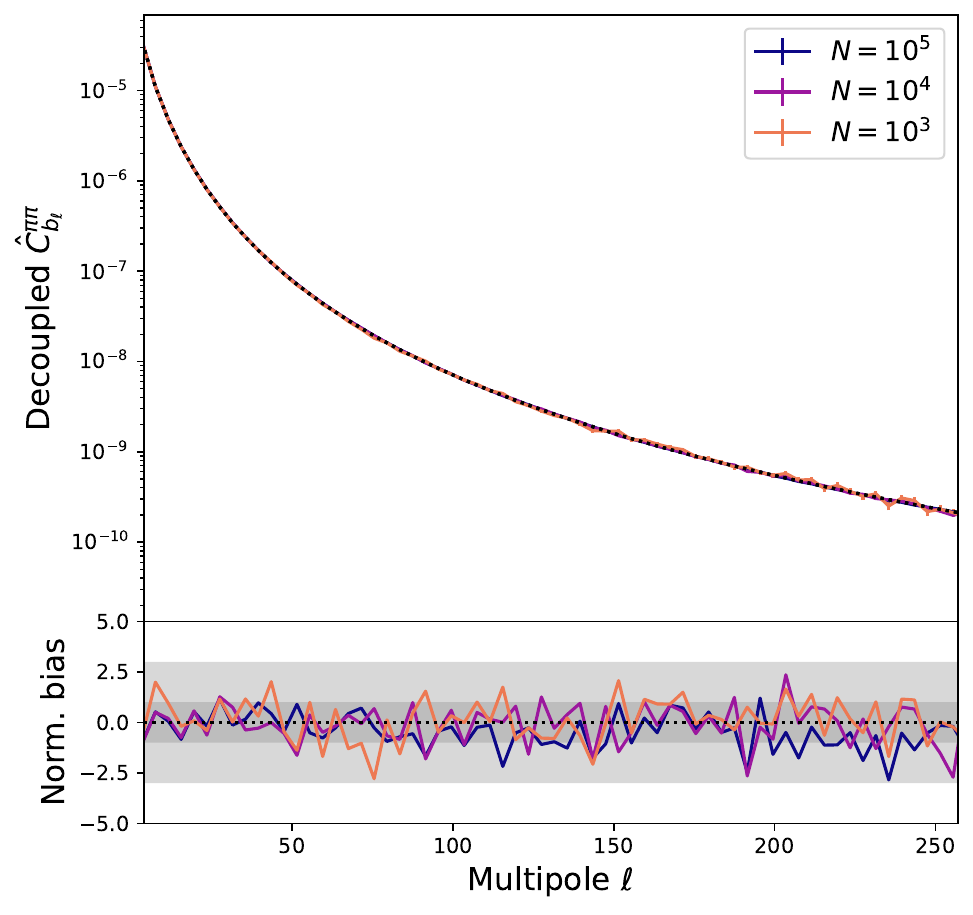}
      \caption{Mode-decoupled catalog$\times$map cross-spectrum for a simulated source catalog and its smooth map-based counterpart at $N_{\rm side}=128$, for catalog sizes of $10^5$, $10^4$, $10^3$. \textit{Upper panel:} mean over 1000 simulations with bars showing the expected standard error on the mean. \textit{Lower panel:} relative bias, in units of the error on the mean.}
      \label{fig:cl_pi_decoupled_sims}
    \end{figure}

    We generate Gaussian realizations of a ``velocity'' field $v_r$ from a known power spectrum
    \begin{equation}
      C_\ell^{vv}=\frac{1}{(\ell+10)^\alpha},
    \end{equation}
    with $\alpha=4$. We then generate an uncorrelated galaxy overdensity map by first constructing a Gaussian field $g(\nv)$ from a power spectrum following the same functional form with $\alpha=3$,\footnote{This is motivated by the fact that, at linear order, the continuity equation makes the velocity power spectrum scale as $P(k)/k$.} and define the overdensity $\delta_g$ from it via a lognormal transformation
    \begin{equation}
        \delta_g(\nv) = \exp\left\{ g(\nv) -\sigma_g^2/2 \right\} -1 \, ,
    \end{equation}
    where $\sigma_g$ is the standard deviation of the map $g$. This ensures that the field is positive-definite ($\delta_g\geq-1$) everywhere. From these maps we obtain the ground truth velocity momentum map via $\pi_g(\nv)=v^g_r(\nv)[1+\delta_g(\nv)]$. To create a catalog-equivalent of $\pi_g$, we construct a galaxy probability map given by $p(\nv)\propto (1+\delta_g(\nv))\,m^g(\nv)$, where $m^g$ is the survey completeness map. For this, we use the selection function of the {\sl Quaia} survey \cite{2306.17749}, shown in Fig. \ref{fig:completeness_mask}. We then generate a Poisson realization of $p(\nv)$ with a desired number density $\bar{n} \equiv\bar{N}/4\pi$. We explore catalogs with source numbers $\bar{N}\in\{10^3,10^4,10^5\}$. We assign to each source $i$ the value $v_r(\nv_i)$ of the velocity field at the exact source position, avoiding any intermediate pixelization effects by direct transformation from the spherical harmonic coefficients $v_{r, \ell m}$ using an irregular-grid inverse SHT \cite{2304.10431}.
    
    Having constructed the base momentum map $\pi_{g,p}$, representing the kSZ map, and the galaxy momentum catalog $\pi_{g,i}$, we calculate their cross-correlation using the methods described in the previous section. Specifically, we construct a standard {\tt NaMaster} {\tt NmtField} object, representing a continuous sky map with a given mask, from $\pi_{g,p}$ masked by the {\sl Quaia} survey selection map. In contrast, the galaxy momentum field is represented by an {\tt NmtFieldCatalogMomentum} object, which uses direct SHTs to obtain the harmonic coefficients of $\pi_{g,i}$ (Eq. \ref{eq:pilm_cat}), and the field's mask is constructed from the completeness map as described in Eq. \ref{eq:alpha_n}. We also validated our implementation of the {\tt NmtFieldCatalogMomentum} field using a random catalog sampled from $m^g$, finding results similar to those presented here.
    The power spectrum between these two fields is then calculated using the standard pseudo-$C_\ell$ steps (see e.g. \cite{1809.09603,2407.21013} for details). To check for estimator bias, we compare against the true input spectrum, obtained from the simulated full-sky momentum map\footnote{To avoid pixelization effects in this comparison, we use this simulation-based estimate of the true spectrum, averaged over all simulations, on large scales, $\ell<150$ while, on small scales $150\leq\ell<2N_{\rm side}$, we employ the analytical result derived in Appendix \ref{app:moments} for the linearized overdensity field $\delta_g\approx e^{-\sigma^2_g/2}(1+g)-1$.}, and convolving with the bandpower window functions associated with the mode-coupling matrix.
    
    We generate 1000 of these simulations using maps defined by the {\tt HEALPix} pixelization scheme \citep{astro-ph/0409513} with resolution $N_{\rm side}=128$. The results of this validation exercise, in terms of the mode-coupled pseudo-$C_\ell$, are shown in Fig. \ref{fig:cl_pi_coupled_sims} for the three different number densities considered here\footnote{Note that, since the mask is the mean source number density, the amplitude of the pseudo-$C_\ell$s (before accounting for mode-coupling due to the mask) scales with the average number density, and hence the three cases explored here (and their theoretical expectations) differ by factors of 10.}. The mode-coupled estimator is unbiased in all cases, with Poisson noise becoming visible by eye at $\ell>100$ in the low-density $\bar{N}=10^3$ case. Fig. \ref{fig:cl_pi_decoupled_sims} shows the mode-decoupled cross-power spectrum of the same set of simulations, compared against the bandpower-convolved ground-truth spectrum. The decoupled estimator remains unbiased for all cases. 

    Although not tested here, in practice, one might also wish to bin a catalog-momentum field with inputs $(\nv_i,\, w_i,\, v^g_{g,i})$ into a pixelized momentum map (see Eq. \ref{eq:mom_pix}) and use it as a map-based field, with the mask constructed from a completeness map as described in Section \ref{sssec:meth.cl.map}. In this case, one must remain mindful of pixelization, aliasing, and other finite-resolution effects (see \cite{2407.21013} for further details).

  \subsection{Comparison against stacking}\label{ssec:res.stack}
    \begin{figure}
      \centering
      \includegraphics[width=0.9\linewidth]{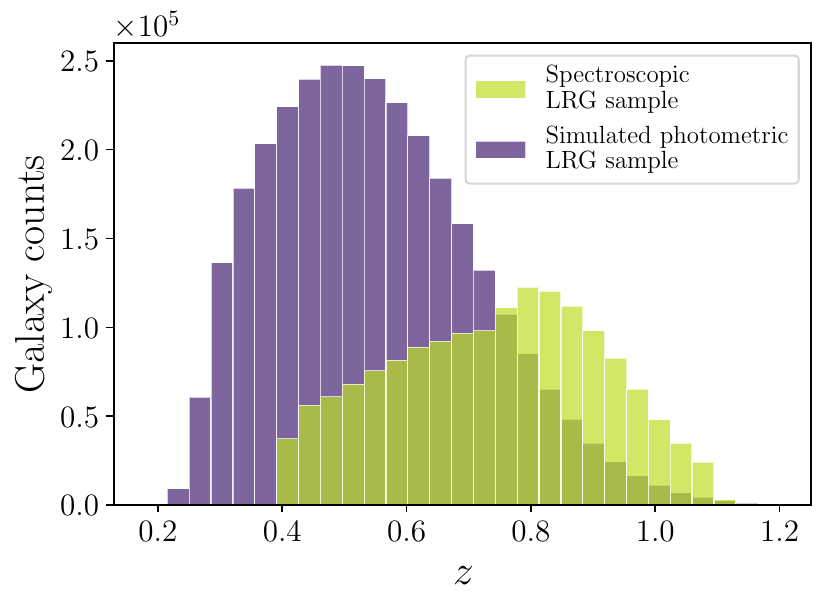}
      \caption{Redshift distribution of the galaxy samples used here. The simulated photometric LRG galaxy sample of \cite{2001.06018}, used in Section \ref{ssec:res.stack}, is shown in purple. The spectroscopic DESI Y1 LRG sample \cite{2208.08515}, used in Section \ref{ssec:res.data}, is shown in green.}\label{fig:SIM_z}
    \end{figure}
    \begin{figure}
      \centering
      \includegraphics[width=\linewidth]{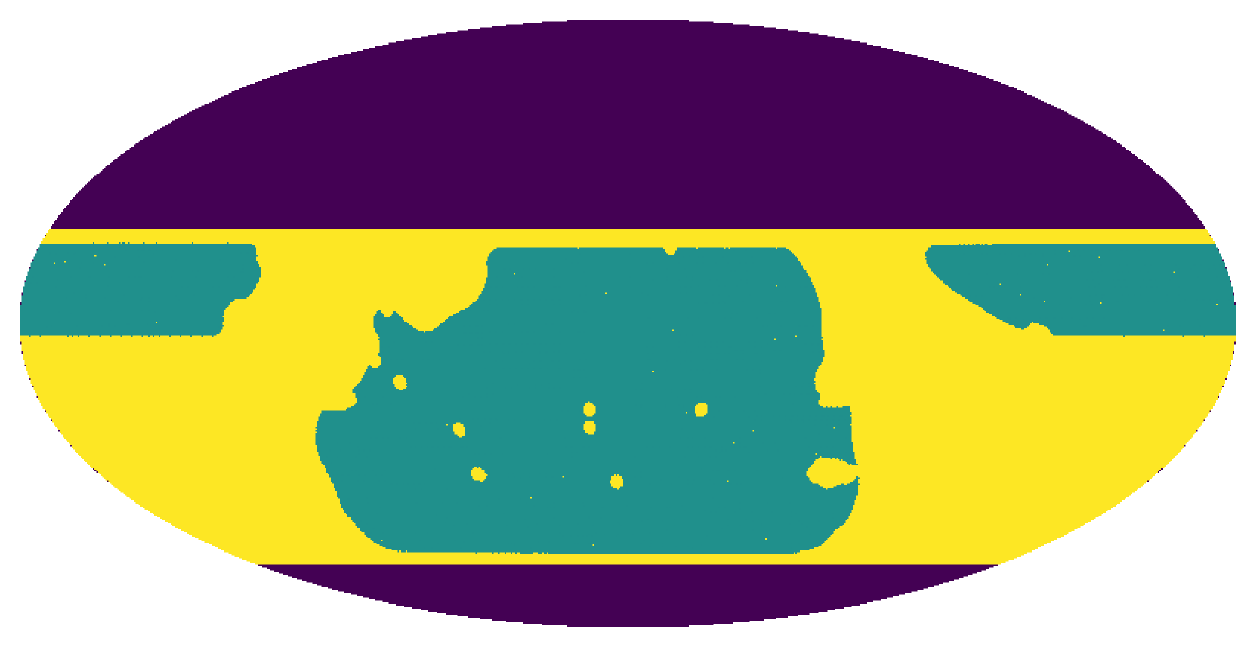}
      \caption{Simulation footprints: the blue region corresponds to the footprint of the galaxy catalog, and is fully enclosed in the footprint of the temperature map (in yellow).}
      \label{fig:SIM_footprint}
    \end{figure}
    Having validated our implementation of the kSZ-galaxy power spectrum estimator, we now apply it to a realistic simulation with three aims:
    \begin{enumerate}[label=(\roman*)]
        \item to verify that the stacking estimator can indeed be recovered from the kSZ-galaxy pseudo-$C_\ell$ following Eq. \ref{eq:cl2stack}; 
        \item to quantify the information content in the power spectrum as opposed to the kSZ stack in scenarios with varying levels of realism; 
        \item to benchmark the computational efficiency of both estimators ($C_\ell$ and stacking).
    \end{enumerate}

    For this exercise, we use a mock galaxy catalog constructed from the \absum simulations \cite{2110.11398, hadzhiyskaHaloLightCone2021}. Specifically, the catalog was designed to mimic the clustering properties of the photometric Luminous Red Galaxy sample (LRG) targeted by the Dark Energy Spectroscopic Instrument (DESI) Legacy Survey. This was constructed using the halo occupation distribution parameters presented in \cite{2001.06018}. We select all galaxies in the sample, covering the redshift range $0.2\lesssim z\lesssim1.1$ and peaking at $z\sim0.5$ (see Fig. \ref{fig:SIM_z}). We also construct a kSZ map by first building a three-dimensional map of the electron overdensity, constructed by scaling the matter overdensity by a baryonic transfer function (see \cite{2504.11794}). The resulting electron density field is scaled by the local value of the radial peculiar velocity, and then integrated along the line of sight to construct a map of the kSZ temperature fluctuation (Eq. \ref{eq:DeltaT_kSZ}). As shown in Fig. \ref{fig:SIM_footprint}, the mock LRG catalog was constrained to a sky footprint mimicking the area covered by the sixth data release of the Atacama Cosmology Telescope collaboration (ACT DR6 -- \cite{2503.14451}), with the kSZ map built over a larger region fully covering this footprint. The map was constructed using equirectangular pixels following the plate carr\'ee pixelization as implemented in {\tt pixell}\footnote{\url{https://github.com/simonsobs/pixell}}, with pixels of size $0.5'$.  For simplicity we weight all galaxies by their true peculiar velocities, both when stacking and when building the galaxy momentum field in the power spectrum estimator, enabling us to set $r_v =1$ in Eq. \ref{eq:defQ}.

    \begin{figure}
      \centering
      \includegraphics[width=\linewidth]{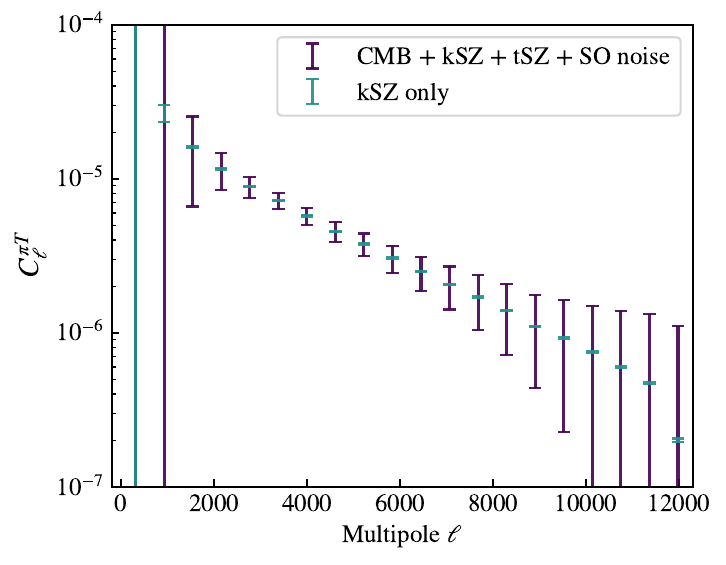} 
      \caption{The decoupled pseudo-$C_\ell$ measured on the kSZ simulation. We measure the $C^{\pi T}_b$ on the pure kSZ realization at a SNR of $373.9$ (turquoise error bars), which drops to $17.7$ (purple error bars) when including the impact of other CMB contributions, foregrounds, and SO-like noise \citep{1808.07445}.
      } \label{fig:SIM_cls}
    \end{figure}

    \begin{figure}
      \centering
      \includegraphics[width=\linewidth]{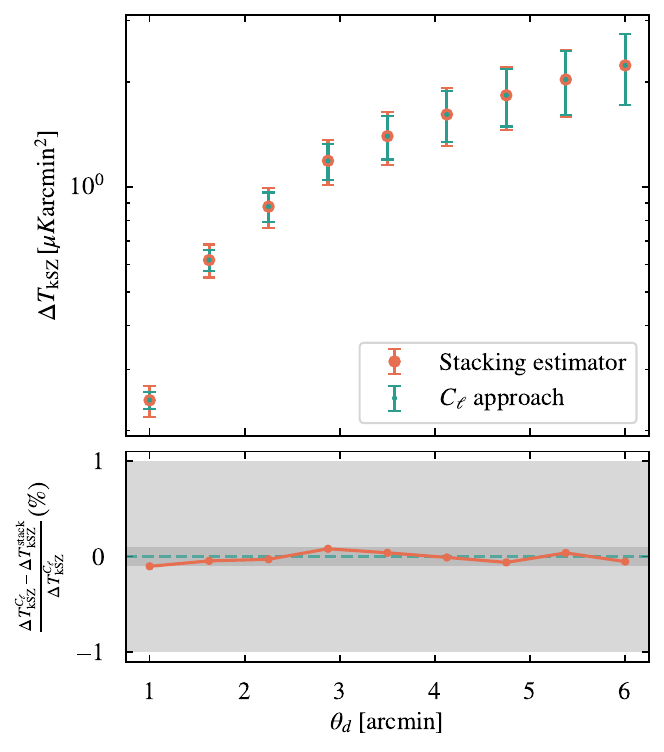}
      \caption{Comparison of the stacking and $C_\ell$ estimators. \textit{Upper panel:} stacked profiles obtained from the stacking estimator used in this work (orange), and from converting the kSZ-galaxy power spectrum $C_\ell^{\pi T}$ (whose decoupled version is shown in Fig. \ref{fig:SIM_cls}) to a stacked profile using Eq. \ref{eq:cl2stack} (turquoise). Note that all error bars have been multiplied by $20$ for better readability. \textit{Lower panel:} relative difference between the stacked profile and its reconstruction from the pseudo-$C_\ell$. We highlight the $\pm 1\%$ and $\pm 0.1\%$ ranges in gray; the residuals lie largely within the $\pm 0.1\%$ range, showing that the stacks can be reconstructed from the pseudo-$C_\ell$ at the estimator level.}\label{fig:SIM_cls_vs_stacks}
    \end{figure}

    \begin{figure*}
      \centering
      \includegraphics[width=0.95\linewidth]{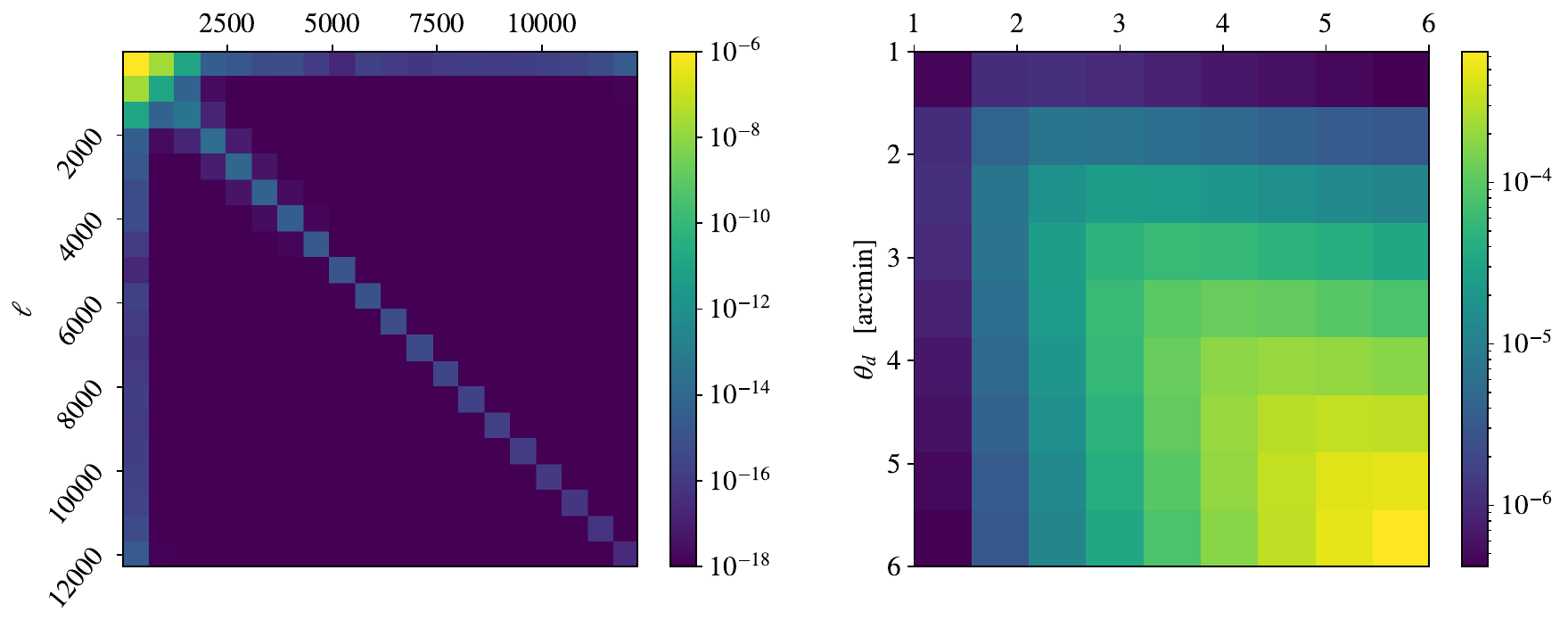}
      \caption{Comparison of the kSZ-galaxy power spectrum covariance $\mathsf C_{bb'}$ (\textit{left}) with the stack covariance $\mathsf C_{\theta \theta'}$ (\textit{right}), both derived from the full $C_\ell^{\pi T}$ covariance, $\mathsf C_{\ell \ell'}$. $\mathsf C_{bb'}$ is mostly diagonal, while $\mathsf C_{\theta \theta'}$ shows large cross-correlations in the off-diagonal entries. Note that the colors are shown using a logarithmic scale, and that we show the absolute value of the $\mathsf C_{b b'}$ covariance in the left panel.}
      \label{fig:SIM_covs}
    \end{figure*}
    Fig. \ref{fig:SIM_cls} shows the mode-decoupled kSZ-galaxy power spectrum computed from the simulation. We bin the multipole range $\ell \in [2; 12{,}281]$ into 20 bandpowers of equal width. The figure includes the statistical uncertainties for this measurement, estimated using the iNKA of \cite{2010.09717}. Note that within this approximation, the covariance only depends on two-point correlators of the fields involved, which can be estimated from the data, including the impact of shot noise. Since the simulated kSZ map contains only the pure kSZ signal, the spectrum is measured at very high precision on all scales, with a total signal to noise ratio (SNR) in the range $\ell\leq12{,}281$ of $\sim 374$. We compute the SNR as \citep{2507.07968}
    \begin{equation}\label{eq:snr}
      \mathrm{SNR} = \sqrt{\mathbf d \mathsf C^{-1} \mathbf d^T - N_\mathrm{dof}} \,, 
    \end{equation}
    where $\mathbf d$ represents the data vector (here, either the $C_b^{\pi T}$s or the stacked profile), $\mathsf C$ its corresponding covariance matrix, and $N_\mathrm{dof}$ is the number of degrees of freedom (here, the length of the data vector). For comparison, we show also the error bars calculated after including the contribution from primary CMB anisotropies, secondary anisotropies and extragalactic foregrounds at 90 GHz, and instrumental noise compatible with the sensitivity that the Simons Observatory (SO) will achieve after 5 years of observations \cite{1808.07445}. These contributions reduce the SNR significantly to 17.7.

    We then use the galaxy-kSZ pseudo-$C_\ell$ to reconstruct the CAP-stacked kSZ signal using Eq. \ref{eq:cl2stack}, and compare the result with the direct stack measurements. For the latter, we compute the stacked kSZ signal using a curved-sky parallelized code. For each galaxy, the code extracts the ``rectangle'' defined by an interval of right ascension and declination that encompasses a spherical cap of radius $\sqrt{2}\theta_d^{\rm max}$, enclosing the outer ring of the largest CAP filter used in the analysis. We consider 
    11 CAP filters with equi-spaced radii in the range $\theta_d\in[1',\,10']$, but will discard $\theta_d > 6'$ to prevent edge effects from contaminating our measurements. As shown in Fig. \ref{fig:SIM_cls_vs_stacks}, we recover the stack measurement with excellent precision (below 0.1\%), showing that all information contained in kSZ stacking is encoded in the $C_\ell$ estimator. We find that this stacking algorithm matches the prediction from the $C_\ell$ better than other algorithms previously used in the literature\footnote{E.g. \url{https://github.com/EmmanuelSchaan/ThumbStack}}. These differences are most prominent (up to $\sim15\%$) on the smaller CAP apertures ($\theta_d\sim1$ arcmin). We ascribe this to interpolation and projection effects, related to the flat-sky projection an re-centering operations commonly applied on the temperature map cutouts for each source \citep{2009.05557}. This is in addition to the impact of the finite pixel size, which distorts the shape of the disk and ring components of the CAP filter. Although pixelization, projection, and interpolation effects may be taken into account by including the associated kernels in the theoretical prediction used to interpret the measurements, the $C_\ell$ approach can avoid these issues altogether.
    
    We can also estimate the covariance matrix of the stacks from the Gaussian covariance of the power spectrum, which we calculate using the iNKA of \citep{2010.09717}. For this we can use Eq. \ref{eq:cl2stack}, according to which, the covariance of the stacks is related to the power spectrum covariance via
    \begin{equation}
      \label{eq:stack_cov_from_cls}
      \mathsf C_{\theta \theta'} = N^2 \sum_{\ell ,\ell'} (2\ell+1)(2\ell'+1) W_\ell (\theta) W_{\ell'}(\theta') \mathsf C_{\ell\ell'}
    \end{equation}
    where $N=(4\pi \, v_\text{rms})^{-1}$. Fig. \ref{fig:SIM_covs} shows the covariances of the mode-decoupled power spectrum bandpowers $\mathsf C_{bb'}$ and the stacks ${\sf C}_{\theta\theta'}$, both derived from the pseudo-$C_\ell$ covariance $\mathsf C_{\ell\ell'}$. As expected, the $C_\ell^{\pi T}$ covariance is mostly diagonal and well-behaved under inversion. The stack covariance on the other hand shows large correlation in the off-diagonal entries, as expected from the cumulative nature of the CAP filter. The error bars associated with this covariance are shown in Fig. \ref{fig:SIM_cls_vs_stacks}, and can be compared there with the uncertainties calculated via bootstrap resampling (estimated from $1000$ samples of the individual stacks with replacement). The Gaussian $C_\ell$ covariance, once converted to its stack form, recovers smaller errors than the bootstrap method at the smallest angular scales, while both estimates agree well at larger separations. A similar result is found for a covariance estimated by shuffling the velocities of different sources in Eq. \ref{eq:stack_def}, effectively canceling out the individual kSZ contributions. The differences found on small scales could be caused by inaccuracies in the bootstrap/reshuffling covariances, or by non-Gaussian contributions to the covariance matrix not captured by the Gaussian power spectrum covariance. We defer a more detailed study of the kSZ power spectrum covariance to future work.
    
    Using this new Gaussian estimate of the stack covariance, we can easily compare the information content in the $C_\ell$ and the stacking estimator. 
    We compute the SNR for each of these using the two covariance matrices used in Fig. \ref{fig:SIM_cls}, which we label ``kSZ-only'' and ``realistic''. To compute the SNR of the stacks, we first convert these two $C_\ell$ covariances to stack covariances using Eq. \ref{eq:stack_cov_from_cls}. For the ``realistic'' case, we find similar SNR values for both $C_\ell$s and stacks of approximately ${\rm SNR} \simeq 18$ (${\rm SNR} = 17.7$ for the $C_\ell$s, and ${\rm SNR} = 18.2$ for the $C_\ell$-derived stack). The fact that both observables recover the same SNR is not entirely surprising, as we should expect the stacks to recover most of the small-scale information, whereas most of the large-scale signal, which the $C_\ell$ could recover more easily, is swamped by the large contribution from the primary CMB. Perhaps more surprisingly, we also find that the stacks and $C_\ell$ recover compatible SNRs in the ``kSZ-only'' scenario, where this large-scale information is more readily available, achieving ${\rm SNR}\simeq374$ in both cases. We therefore conclude that the level of information loss inherent in compressing the $C_\ell$ data into CAP-filtered stacks is largely negligible.

    We note two subtleties when comparing the SNR of both of these estimators. First, binning the power spectrum into bandpowers may incur a small loss of information. Secondly, inverting the stack covariance can be numerically unstable, due to its large off-diagonal entries. The combination of these two effects can lead to a marginally larger SNR for the stacks than the $C_\ell$s in some cases, even though the former can be constructed from the latter, as we have shown. We also estimated the SNR of the pseudo-$C_\ell$ before binning it into mode-decoupled bandpowers\footnote{We use singular-value decomposition in this case, since the unbinned inverse covariance is numerically unstable.}. In this case we find an SNR that is marginally larger than either the stacks or the $C_\ell$ bandpowers (e.g. ${\rm SNR}=378$ in the ``kSZ-only'' case), confirming that a similar amount of information is lost in binning or convolving with the CAP filters.

    \begin{figure}
      \centering
      \includegraphics[width=0.95\linewidth]{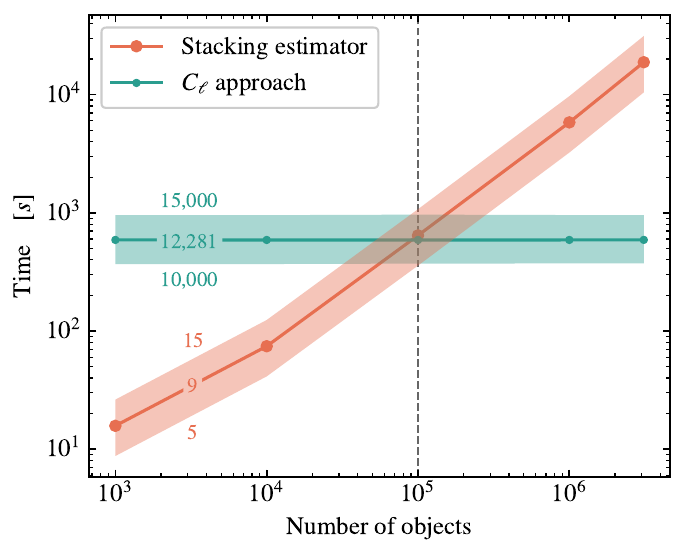}
      \caption{Computational efficiency of both estimators as a function of the number of objects included in the catalog. The stacking estimator runtime performance (orange line) exhibits a linear trend, while the $C_\ell$ estimator performance (turquoise line) remains roughly constant with an increasing number of objects. Note that the performance of the stacking estimator depends on the number of angular separation values at which the profile is computed (here, the 9 values of $\theta_d$ from Fig. \ref{fig:SIM_cls_vs_stacks}), and that of the $C_\ell$ estimator depends on the $\ell_\text{max}$ up until which the power spectrum is computed (here, $\ell_\text{max} = 12{,}281$). For reference, we also show, as an orange band, the expected stacking performance for 5 angular separations (lower bound) and 15 angular separations (upper bound), and, as a turquoise band, the $C_\ell$ performance for $\ell_\text{max} = 10{,}000$ (lower bound) and $\ell_\text{max} = 15{,}000$ (upper bound). For simplicity, times are provided for runs on a single computer core, but both estimators can be fully parallelized.}
      \label{fig:SIM_runtime}
    \end{figure}

    Finally, we compare the computational efficiency of both the stacking and $C_\ell$ estimators. We repeat measurements for different sample sizes $N$ ($N=10^3$, $10^4$, $10^5$, $10^6$ and the full simulated catalog, i.e. $N=3.6 \times 10^6$ objects). For the stacking estimator, we only compute the individual profiles at the 9 angular separations shown in Fig. \ref{fig:SIM_cls_vs_stacks}; for the $C_\ell$ estimator, we compute the pseudo-$C_\ell$ up to the same $\ell_\mathrm{max} = 12{,}281$, but we do not include the decoupling and binning operations in the runtime performance. This is because these operations are discretionary, need only be performed once for a given set of sky masks, and can become a large runtime contribution as they depend on the binning scheme used\footnote{For reference, for the power spectra shown in Fig. \ref{fig:SIM_cls}, the mode-coupling matrix takes about 1.7 core hours to compute up to $\ell_\mathrm{max} = 12{,}281$ -- or about 10 minutes on a 10-core machine.}. The result of this comparison is shown in Fig. \ref{fig:SIM_runtime}. For both estimators, we only present runtime that pertains to the actual measurement (the loading of data products, common in both cases, is not included). The stacking estimator's runtime evidently increases linearly with the number of objects in the sample, and depends on the radial binning -- as illustrated by the orange band in the figure. The $C_\ell$ estimator, on the other hand, shows a roughly constant runtime performance, nearly independent of the size of the catalog and scaling mainly with $\ell_{\rm max}$ (as shown by the turquoise band). For the radial binning used in this work, the $C_\ell$ estimator outperforms the stacking estimator at around a sample size of $N = 10^5$ objects, corresponding to a rather small galaxy catalog. For the size of the simulated catalog, the $C_\ell$ estimator is more than 30 times faster than the stacking estimator. It must be noted that, although the times reported in Fig. \ref{fig:SIM_runtime} correspond to runs on a single core, both tasks (pseudo-$C_\ell$ and stacking) are easily and fully parallelizable.
    \begin{figure}
      \centering
      \includegraphics[width=\linewidth]{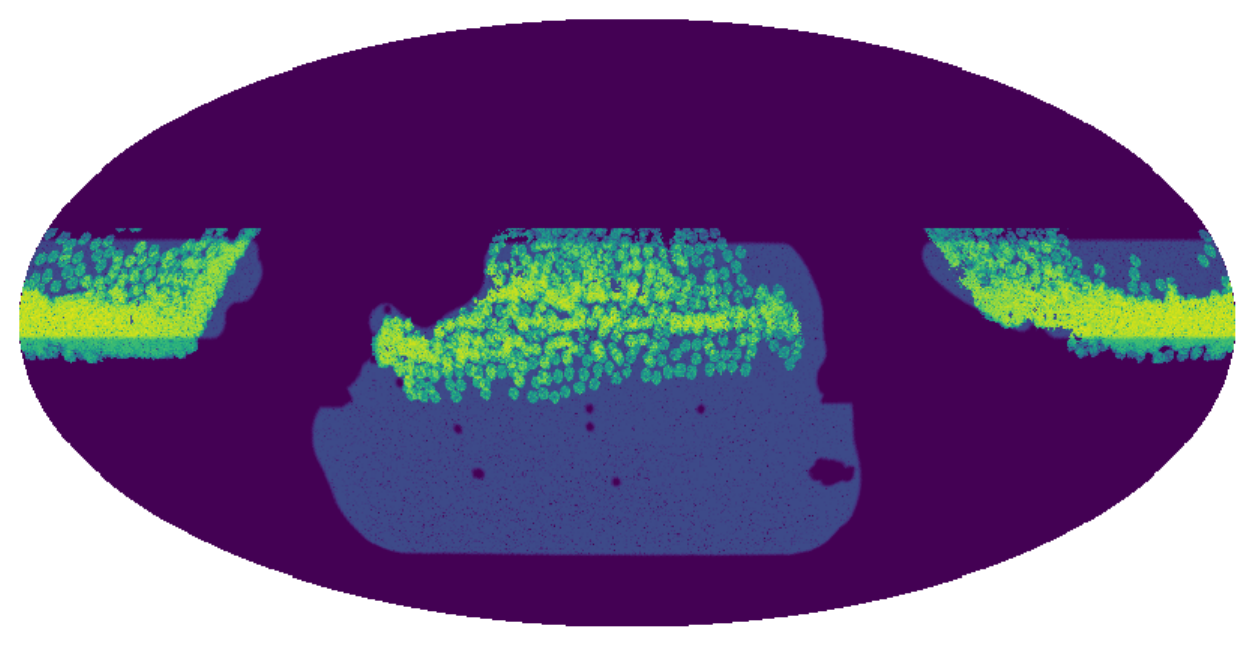}
      \caption{Datasets footprints: the dark blue footprint corresponds to the apodized ACT mask, which varies between 0 and 1. The yellow-green footprint is that of the DESI Y1 spectroscopic LRG catalog; variations in this mask track the number density of galaxies observed. We limit the galaxy catalog to those sources overlapping with the ACT mask.}
      \label{fig:DATA_footprint}
    \end{figure}
    
    \begin{figure*}[]
      \centering
      \subfigure{\includegraphics[width=0.46\textwidth]{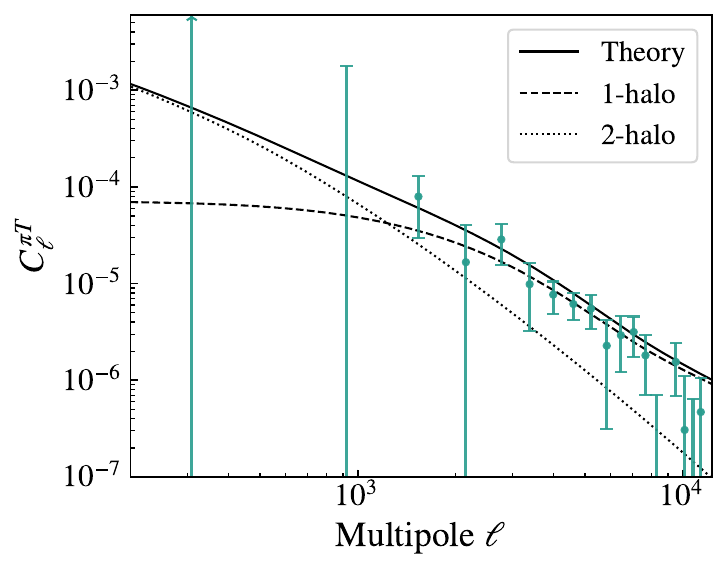}} 
      \quad\quad
      \subfigure{\includegraphics[width=0.46\textwidth]{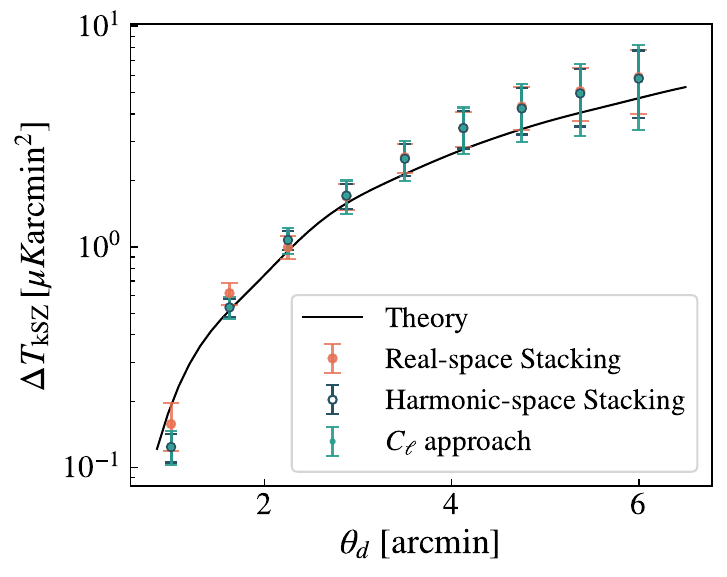}} 
      \caption{\textit{Left panel}: measurement of the kSZ-galaxy power spectrum from ACT DR6 and the DESI Y1 LRG sample (turquoise data points) with a theoretical prediction for the baryonic model of \cite{postaXyInsightsGas2024} with the parameters listed in Eq. \ref{eq:bcm_params}, and a correlation coefficient of $r_v=0.64$. \textit{Right panel}: comparison of the corresponding stacked radial profiles and theory line with other stacking methods. In turquoise, we show the stack recovered from the coupled kSZ power spectrum; in dark blue, the ``harmonic stack'' estimate as described in the main text, and in orange the standard stacking estimator. Bootstrap errors are shown for the standard and harmonic stacking estimators, and Gaussian errors (Eq. \ref{eq:stack_cov_from_cls}) for the $C_\ell$-derived profile.}\label{fig:DATA_cls_vs_stacks}
    \end{figure*}

\subsection{Demonstration on real data}\label{ssec:res.data}
  As a first demonstration of the kSZ power spectrum estimator on real data, we apply it to the cross-correlation between the component-separated ``CMB+kSZ'' ACT DR6 temperature map, presented in \cite{2307.01258}, and the DESI Y1 LRG spectroscopic galaxy sample \cite{2208.08515}. LRGs are an ideal sample for kSZ measurements due to their higher clustering bias and optical depth, and low satellite fraction. The latter makes them less sensitive to observational systematics such as off-centering \cite{guachallaBacklightingExtendedGas2025,2509.18732}. Most importantly, the ACT $\times$ DESI LRG Y1 kSZ stacked profile has already been measured \cite{guachallaBacklightingExtendedGas2025}, allowing us to compare our results against those found in the literature. The redshift distribution of the DESI LRG Y1 sample is shown in Fig. \ref{fig:SIM_z} (green histogram, spanning $0.4 \lesssim z \lesssim 1.1$), and the sky masks of both datasets are shown in Fig. \ref{fig:DATA_footprint}. Note that, for our fiducial measurements, we select only galaxies in the LRG sample that fall within the ACT DR6 footprint. We use reconstructed velocities computed using the \texttt{pyrecon} package\footnote{\url{https://github.com/cosmodesi/pyrecon}}   \cite{paillasOptimalReconstructionBaryon2025}.

  We use the same binning scheme employed in Section \ref{ssec:res.stack}, both for the multipole bandpowers and angular separations. The kSZ-galaxy power spectrum measurement is shown in the left panel of Fig. \ref{fig:DATA_cls_vs_stacks}. We detect a clear correlation between the kSZ map and the galaxy momentum field, with a total SNR of $8.7$ (estimated via Eq. \ref{eq:snr}), using angular scales $\ell\le 12{,}281$. The stacking measurements from the same data are shown in the right panel of the same figure (orange points with error bars), with the stack inferred from the $C_\ell$ measurement using Eq. \ref{eq:cl2stack}\footnote{Note that, as we use the same reconstructed velocities in both measurements, we assume $r_v=1$ in this conversion.} shown in turquoise.

  The two approaches agree well at large angular separation and, despite some differences, are still consistent within the errors at small angular separations. As we described in Section \ref{ssec:res.stack}, these differences are expected due to the impact of finite pixel effects, which are exacerbated in the case of real data by the presence of small-scale noise, leading to larger variations between nearby pixels. We verified that the small differences found in Section \ref{ssec:res.stack} grow significantly when adding noise to the simulated kSZ map. To further verify that the differences observed are due to pixelization effects, we implement an alternative stacking method, free from such effects, which we label ``harmonic-space stacking''. In this case, the CAP-filtered signal at the position of a given source is calculated by multiplying the harmonic transform of the CMB map by the SHT of the CAP filter window function, and then evaluating this CAP-convolved map at the exact source position via an irregular-grid inverse SHT (as done in Section \ref{ssec:res.val} to avoid pixelization effects). The result of this method is shown as dark blue points in Fig. \ref{fig:DATA_cls_vs_stacks}, and agree with the $C_\ell$-derived signal at very high precision.

  As in Section \ref{ssec:res.stack}, the errors shown in the right panel of Fig. \ref{fig:DATA_cls_vs_stacks} were obtained via bootstrapping the individual profiles for both real- and harmonic-space stacks, and by converting the Gaussian power spectrum covariance using Eq. \ref{eq:stack_cov_from_cls} for the $C_\ell$-derived stack. Interestingly, this time the analytical Gaussian errors are larger than their empirical (bootstrap) counterparts. This partly accounts for the slightly lower SNR at which the $C_\ell$-derived stack (equivalently, the kSZ-galaxy power spectrum) is detected in this work, compared to the previous results of \cite{guachallaBacklightingExtendedGas2025} who detect the LRG-kSZ correlation at the $\sim10\sigma$ level. We illustrate this point using our hybrid harmonic stacking estimator, which matches both the $C_\ell$ approach in terms of stack amplitude and the real-space stacking in terms of covariance: the harmonic stack, together with its bootstrap covariance, is detected at $\text{SNR}=11.2$, but its SNR drops back to the same SNR value as the power spectrum's ($\text{SNR}=8.7$) when using instead the Gaussian power spectrum covariance estimate. We leave a more thorough study of the kSZ power spectrum covariance, and the appropriateness of bootstrap uncertainties, for future work.

  In the standard stacking measurements, a subset of the galaxies are further excluded to avoid those that lie too close to point sources or to the mask edges, in order to mitigate biases to the measurements caused by survey geometry effects. One of the advantages of the pseudo-$C_\ell$ estimator is its immunity to these effects. However, in general, limiting the sample to the overlap between the two fields being correlated may lead to an enhancement in the measured cross-correlation, since the map fluctuations far away from the overlapping region could act as a source of noise in harmonic space. It is therefore interesting to consider the impact in our measurements of removing the galaxies in the LRG sample that lie outside of the ACT footprint. Repeating our analysis using the full LRG sample shown in Fig. \ref{fig:DATA_footprint}, we find that the total SNR drops to $\text{SNR}=8.5$. This small change is caused by a few-percent reduction in the amplitude of the measured spectrum, which in turn leads to smaller errors (as the Gaussian covariance estimate depends on the power spectra themselves). This change in amplitude could be a simple statistical fluctuation, or a sign of potential large-scale systematics in the galaxy distribution, leading to a spatial modulation in the density or reconstructed galaxy velocities.

  In Fig. \ref{fig:DATA_cls_vs_stacks}, we plot theoretical predictions for the kSZ power spectrum (left panel) and stacks (right panel), based on the framework described in \cite{2509.18732}. For simplicity, we neglect the subdominant contributions presented in \cite{2509.18732}, such as the cross-term and the connected trispectrum. The prediction is based on a baryonic correction model \cite{schneider2015}, specifically that described in \cite{postaXyInsightsGas2024}, in which the electron density profile is described by 5 parameters: $\log_{10}M_c$ sets the mass scale at which half of the gas content has been ejected by AGN outflows. $\beta$ characterizes the mass dependence of the ejected fraction. $\eta_b$ quantifies the radius out to which the ejected gas is expelled. The polytropic index $\Gamma$ determines the density profile of the bound gas. Finally, $A_*$ quantifies the fraction of the gas converted into stars (see \cite{postaXyInsightsGas2024} for further details). The amplitude of the theoretical prediction is also modified by the correlation coefficient $r_v$ between the reconstructed and true velocities. We use $r_v=0.64$ for the LRG sample used here \cite{guachallaBacklightingExtendedGas2025}. The theoretical prediction shown in Fig. \ref{fig:DATA_cls_vs_stacks} corresponds to a model with parameters
  \begin{equation}\label{eq:bcm_params}
    \{\log_{10}M_{\rm c},\eta_{\rm b},\beta,A_*,\Gamma\}=\{14.9,0.6,0.6,0.03,1.16\},
  \end{equation}
  roughly compatible with the model found in \cite{postaXyInsightsGas2024} to describe the correlation between weak lensing, thermal SZ, and X-ray data. Although we have not attempted to a robust analysis of our measurements within this model, the theory prediction shows a remarkable level of agreement with the kSZ measurements presented here. Moreover, Fig. \ref{fig:DATA_cls_vs_stacks} also allows us to assess the importance of secondary contributions to the kSZ-galaxy power spectrum, such as the 2-halo term (shown as a dotted line in the figure). Where this term becomes relevant ($1000\lesssim\ell \lesssim 3500$), the error on the measurement is roughly comparable to the size of its contribution, suggesting that the 2-halo term may not be neglected in future modeling of the kSZ-galaxy power spectrum (as pointed out in \cite{2509.18732}). We leave a more thorough theoretical exploration of the kSZ power spectrum measurements in combination with other gas probes for future work.
    
\section{Conclusions}\label{sec:conc}
  This paper presents a new angular power spectrum approach to measuring the kSZ effect on projected fields; we show that this approach is mathematically equivalent to the more traditional kSZ ``stacking estimator'', while offering a number of advantages. This work is part of a wider effort towards simplifying the inclusion of baryonic probes into cosmological analyses, from early work showing equivalence between real-space kSZ estimators and the $ggT$ bispectrum \cite{smithKSZTomographyBispectrum2018}, to current efforts to express the kSZ effect as a well-understood power spectrum measurement \cite{hillKinematicSunyaevZeldovichEffect2016, quinprep, hadzhiyskainprep}. 

  The estimator presented here consists in taking the cross-power spectrum of a galaxy momentum density field $\pi_g$, constructed from individual galaxy positions and velocities, with the CMB temperature map $T$. By re-using the existing infrastructure for power spectrum measurements, including catalog-based methods to bypass pixelization effects and aliasing, the same information can be extracted from the data at a much lower computational cost. The resulting measurements are robust to the effects of sky masks (e.g. mode-coupling), and come with well-tested covariance estimates. As such, the kSZ effect can be seamlessly incorporated into large-scale structure analyses that use the same power spectrum framework, and the cross-covariances between probes can be obtained easily and reliably.

  We validate our implementation of this  estimator using a toy model in which we simulate fields and catalogs from known power spectra, and compare their cross-correlation to the estimator outputs. Having validated the estimator, we apply it to both simulated and real data. We find that the $C_\ell$ approach, once converted to a stacked profile, reproduces the stacking approach, recovering all information present in the latter. Although we find small differences between both at small scales, we can trace these to the impact of finite pixel size, interpolation, and flat-sky projection, effects that the power spectrum estimator avoids altogether. We showed also that the power spectrum estimator offers a more scalable computational performance than stacking for large catalogs.

  Finally, we applied this estimator to real data for the first time, specifically CMB maps from ACT DR6, and the DESI Y1 spectroscopic LRG sample. We show that the kSZ-galaxy power spectrum can be detected at high significance (SNR $\sim 9$), and that the result is compatible with stacking measurements at the estimator level once finite-resolution effects are taken into account. A tentative and qualitative comparison of this measurement against theoretical predictions using a simple parametrization of the gas distribution in halos (using the formalism described in \cite{2509.18732}) shows a reasonable agreement with the physical models favored by the analysis of other independent gas probes (specifically tSZ and X-ray data in \cite{2412.12081}).

  In summary, the estimator presented here is ready for deployment on real data, and allows for the inclusion of kSZ cross-correlation in multi-tracer cosmological analyses in a seamless and robust manner. To ensure a reliable accounting and propagation of all statistical uncertainties, a more detailed study of the kSZ covariance (in particular contributions beyond those readily included in the standard pseudo-$C_\ell$ Gaussian covariance framework) is necessary, a problem that we have not addressed here. The work presented here could also be generalized to reproduce the information in other stacking estimators in a pseudo-$C_\ell$ framework. This would be particularly interesting in the context of ``directional'' stacking analyses carried out in e.g. the search for signals aligned with cosmic filaments \cite{2412.03631}, the moving-lens effect \cite{2305.15462,2408.16055}, and general dipoles correlated with the transverse velocity field \cite{2504.02525}. This will be the subject of future work.

\section*{Acknowledgments}
  When the work presented here was at an advanced stage, we learnt of similar work being developed by Qu et al (2026) \cite{quinprep} and Hadzhiyska et al (2026) \cite{hadzhiyskainprep}. We thank Emmanuel Schaan, and the other authors of these works for useful discussions. We also thank Adrien La Posta and Andrina Nicola for useful comments. LH is supported by a Hintze studentship, which is funded through the Hintze Family Charitable Foundation. KW acknowledges support from the Science and Technology Facilities Council (STFC) under grant ST/X006344/1. DA acknowledges support from STFC and the Beecroft Trust. AW is supported by a Science and Technology Facilities Council (STFC) studentship. We made extensive use of computational resources at the University of Oxford Department of Physics, funded by the John Fell Oxford University Press Research Fund. 

\onecolumngrid

\appendix

\section{The power spectrum of $(1+\delta)\,f$}\label{app:moments}
  \newcommand{\wtj}[6]{\left(\begin{array}{ccc} #1 & #2 & #3\\#4 & #5 & #6\end{array} \right)}
  Let $\pi(\nv)=(1+\delta(\nv))f(\nv)$, with $\delta(\nv)$ the overdensity field, and $f(\nv)$ another field (e.g. the radial velocity, in the case of a momentum density field). The harmonic coefficients of $\pi$ are then:
  \begin{equation}
    \pi_{\ell m}=f_{\ell m}+\sum_{\ell'm',\ell''m''}{\cal G}^{\ell\ell'\ell''}_{mm'm''}\delta_{\ell'm'}f_{\ell''m''},\hspace{12pt}{\rm where}\hspace{12pt}
    {\cal G}^{\ell\ell'\ell''}_{mm'm''}\equiv\int d\nv\,Y^*_{\ell m}(\nv)Y_{\ell'm'}(\nv)Y_{\ell''m''}(\nv).
  \end{equation}

  The power spectrum of $\pi$ is then
  \begin{align}
    C_\ell^{\pi\pi}
    &=C_\ell^{ff}+\sum_{\{\ell m\}_{1,2,3,4}}{\cal G}^{\ell\ell_1\ell_2}_{mm_1m_2}{\cal G}^{\ell\ell_3\ell_4}_{mm_3m_4}\,\langle \delta_{\ell_1m_1}f_{\ell_2m_2}\delta_{\ell_3m_3}^*f_{\ell_4m_4}^*\rangle\\
    &=C_\ell^{ff}+\sum_{\ell_1\ell_2}\left[\sum_{m_1m_2}\left({\cal G}^{\ell\ell_1\ell_2}_{mm_1m_2}\right)^2\right]\left(C_{\ell_1}^{\delta\delta}C_{\ell_2}^{ff}+C_{\ell_1}^{\delta f}C_{\ell_2}^{\delta f}\right)
  \end{align}
  where, in the second line, we have assumed that all fields are Gaussian. Finally, expressing ${\cal G}^{\ell\ell'\ell''}_{mm'm''}$ in terms of Wigner $3j$ symbols, and using their orthogonality properties, we obtain the final result
  \begin{align}
    C_\ell^{\pi\pi}
    &=C_\ell^{ff}+\sum_{\ell_1\ell_2}\frac{(2\ell_1+1)(2\ell_2+1)}{4\pi}\wtj{\ell}{\ell_1}{\ell_2}{0}{0}{0}^2\left(C_{\ell_1}^{\delta\delta}C_{\ell_2}^{ff}+C_{\ell_1}^{\delta f}C_{\ell_2}^{\delta f}\right).
  \end{align}

\twocolumngrid

\bibliography{main}

\end{document}